\documentclass[letterpaper,twocolumn,aps,prl,showpacs,floatfix,superscriptaddress]{revtex4-1}
\usepackage{bm}
\usepackage{amsmath}
\usepackage{graphicx}
\usepackage{epstopdf }
\usepackage[version=3]{mhchem}
\usepackage[colorlinks=true,linkcolor=blue,citecolor=blue,urlcolor=blue]{hyperref}
\usepackage{natbib}
\newcommand{\eat}[1]{}

\def\br{\boldsymbol{r}}
\def\barj{\bar{\jmath}}

\begin{document}

\title{High-energy electronic excitations in Sr$_2$IrO$_4$ observed by Raman scattering}
\author{Jhih-An Yang}
\affiliation{Department of Physics, University of Colorado at Boulder, Boulder, CO 80309}
\author{Yi-Ping Huang}
\affiliation{Department of Physics, University of Colorado at Boulder, Boulder, CO 80309}
\author{Michael Hermele}
\affiliation{Department of Physics, University of Colorado at Boulder, Boulder, CO 80309}
\author{Tongfei Qi}
\affiliation{Department of Physics and Astronomy, University of Kentucky, Lexington, Kentucky 40506}
\author{Gang Cao}
\affiliation{Department of Physics and Astronomy, University of Kentucky, Lexington, Kentucky 40506}
\author{Dmitry Reznik}
\affiliation{Department of Physics, University of Colorado at Boulder, Boulder, CO 80309}

\date{\today}

\begin{abstract}
Spin-orbit interaction in \ce{Sr2IrO4} leads to the realization of the $J_{\mathrm{eff}}$ = 1/2 state and also induces an insulating behavior. Using large-shift Raman spectroscopy, we found two high-energy excitations of the d-shell multiplet at 690 meV and 680 meV with $A_{1g}$ and $B_{1g}$ symmetry respectively. As temperature decreases, the $A_{1g}$ and $B_{1g}$ peaks narrow, and the $A_{1g}$ peak shifts to higher energy while the energy of the $B_{1g}$ peak remains the same. When 25$\%$ of Ir is substituted with Rh the $A_{1g}$ peak softens by 10$\%$ but the $B_{1g}$ peak does not. We show that both pseudospin-flip and non-pseudosin-flip dd electronic transitions are Raman active, but only the latter are observed. Our experiments and analysis place significant new constraints on the possible electronic structure of \ce{Sr2IrO4}.

\end{abstract}
\maketitle

\section{I. Introduction}

In $5d$ transition-metal oxides (TMOs), the $d$ electrons are extended in real space, which should lead to a large bandwidth (W) and a reduced on-site Coulomb correlation (U). As a result, $5d$ TMOs are expected to be metals as opposed to $3d$ TMOs, which are Mott insulators. However, spin-orbit coupling (SOC) in $5d$ TMOs is an order of magnitude larger than in $3d$ TMOs due to the large atomic number.  Several iridium oxides including \ce{Sr2IrO4} are insulators because of SOC \cite{PhysRevB.49.9198,PhysRevB.57.R11039,PhysRevB.82.064412,PhysRevB.83.155118,nature_Pesin}. It has been proposed that the interplay between SOC, Coulomb correlation, crystal field splitting, and inter-site hopping can lead to unconventional electronic states for the $5d$ TMOs \cite{Jackeli2009}.

\ce{Sr2IrO4} attracted attention as the first realization of a SOC-induced insulator \cite{PhysRevLett.101.076402,Kim06032009}. The spin-orbit interaction (SOI) $\lambda\sim0.4$ eV in \ce{Sr2IrO4} splits the $t_{2g}$ states into bands with $J_{\mathrm{eff}}$ = 1/2 and $J_{\mathrm{eff}}$ = 3/2. Ir$^{4+}$ ($5d^{5}$) provides five electrons, and thus the $J_{\mathrm{eff}}$ = 3/2 state is fully occupied and the $J_{\mathrm{eff}}$ = 1/2 state is half filled. The narrow $J_{\mathrm{eff}}$ = 1/2 band then splits into the lower Hubbard band (LHB) and the upper Hubbard band (UHB) due to on-site repulsion, resulting an insulating state \cite{PhysRevLett.101.076402,PhysRevLett.101.226402,Kim06032009,PhysRevLett.105.216410,PhysRevLett.107.266404}. Optical conductivity, angle-resolved photoemission spectroscopy, x-ray absorption spectroscopy \cite{PhysRevLett.101.076402,Moon2006}, and resonant inelastic x-ray scattering (RIXS) \cite{PhysRevB.83.115121} results are consistent with this scenario, though it has also been argued that \ce{Sr2IrO4} is a magnetically-ordered Slater insulator \cite{PhysRevLett.108.086403,PhysRevB.86.035128,PhysRevB.89.121111}. 

Earlier Raman scattering experiments revealed excitations in undoped cuprates around 1.5 eV that were assigned to d-d excitations \cite{PhysRevB.51.6617,PhysRevB.53.886,PhysRevLett.71.3709}. At low temperatures these measurements appeared as closely-spaced broad peaks with the onset just below the gap. Here we report similar excitations in \ce{Sr2IrO4} at lower energies ($\sim$0.7 eV). A peak at the same energy has been previously observed by RIXS \cite{PhysRevLett.108.177003,nature_Kim}. Using Raman scattering measurements with polarization analysis and superior resolution we established that the RIXS feature consists of electronic excitations of  $A_{1g}$ and $B_{1g}$ symmetry. Each contribution (especially  $A_{1g}$) is surprisingly narrow in energy, and the two contributions have different temperature and Rh-doping dependences. Our results put strict constraints on any future theory of electronic structure of  \ce{Sr2IrO4}.

\section{II. experimental details}

High-quality single crystal samples of doped and undoped \ce{Sr2IrO4} were synthesized as described elsewhere \cite{PhysRevB.57.R11039}. \ce{Sr2IrO4} has a tetragonal crystal structure, which belongs to the space group $I4_{1}/acd$ \cite{PhysRevB.49.9198}. The \ce{IrO6} octahedra are rotated about the c axis by $\sim11^{\circ}$. 

Different laser lines (457.9-560 nm) from Kr and Ar ion lasers as well as a 532 nm solid state laser were used with a power of 15 mW for the measurements at room temperature and base temperature. Temperature dependence measurements were performed with 2 mW to minimize laser heating. All experiments were performed on a McPherson custom triple spectrometer equipped with a cooled charge-coupled device (CCD) detector. It was configured in a subtractive mode with 50 grooves per mm gratings in the filter stage and 150 g/mm in the spectrometer stage, which gave the resolution of $ \sim $8 meV. The sample was mounted in a top-loading closed-cycle refrigerator. The entrance slit of the spectrometer was opened to 0.5 mm to avoid chromatic aberrations of the collecting optics. The data were corrected for the spectral response of the equipment using a calibrated lamp with a broad spectrum. 


The configuration xx/xy denotes that the incident laser polarization is parallel to the primitive cell in-plane crystal axes (a and b), and the scattered light polarization is parallel/perpendicular to the incident laser polarization respectively. The x' and y' directions are rotated $45^\circ$ in the ab plane with respect to a and b. xx, xy, x'x', and x'y' polarization geometries measure $A_{1g}+B_{1g}$, $A_{2g}+B_{2g}$, $A_{1g}+B_{2g}$, and $A_{2g}+B_{1g}$ symmetry components respectively in the $D_{4h}$ point group to which the \ce{IrO6} octahedra belong. Although the full site symmetry is lowered to $C_{4h}$ by the rotation of the octahedra, we will first discuss the results in terms of the representations of $D_{4h}$ site symmetry and then show that the symmetry lowering does not have a measurable effect.

In addition to the Raman peaks, a broad background appears as was reported in Ref. \cite{PhysRevB.85.195148}. We found that it depends strongly on surface preparation and on the part of the sample that is being probed. The lowest background was obtained on samples that were cleaved in high vacuum before being transferred to the cryostat. Based on our investigation, it appears that this background is mostly an artifact of surface degradation although we cannot rule out a component that is intrinsic. 

\section{III. RESULTS AND DISCUSSION}
\subsection{A. Experimental electronic excitations}
\begin{figure}[t]
\includegraphics[width=0.4\textwidth]{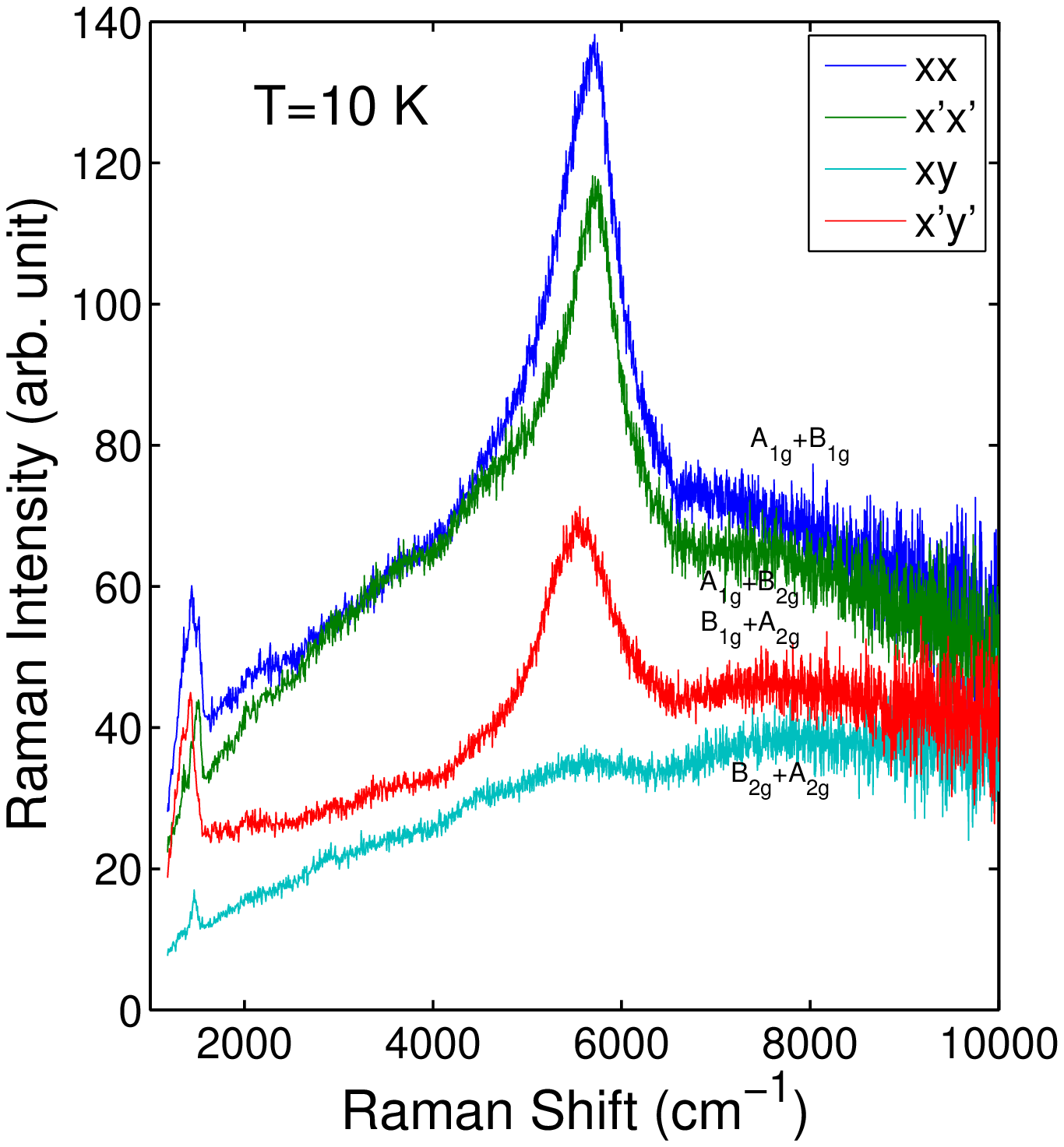} 
\caption{Raman spectra with incident laser wavelength 457.9 nm in four different scattering configurations measured at 10 K. Broad peaks around 5600 $cm^{-1}$ are electronic (see text) and peaks around 1400 $cm^{-1}$ are two-phonon scattering \cite{PhysRevB.85.195148}.  \label{458nm}}
\end{figure}

Electronic Raman scattering from \ce{Sr2IrO4} is dominated by strong peaks near 5600 $cm^{-1}$  that appear in xx, x'x', and x'y' geometries, but not in the xy geometry beyond what is expected from imperfect polarization analysis (Fig. \ref{458nm}) i.e. these peaks exist in the $A_{1g}$ and $B_{1g}$ symmetries but not in $B_{2g}$ or $A_{2g}$ symmetry. The $B_{1g}$ peak appears at 5500 $cm^{-1}$, and the $A_{1g}$ peak is at 5600 $cm^{-1}$.

\begin{figure*}[t]
\includegraphics[width=1\textwidth]{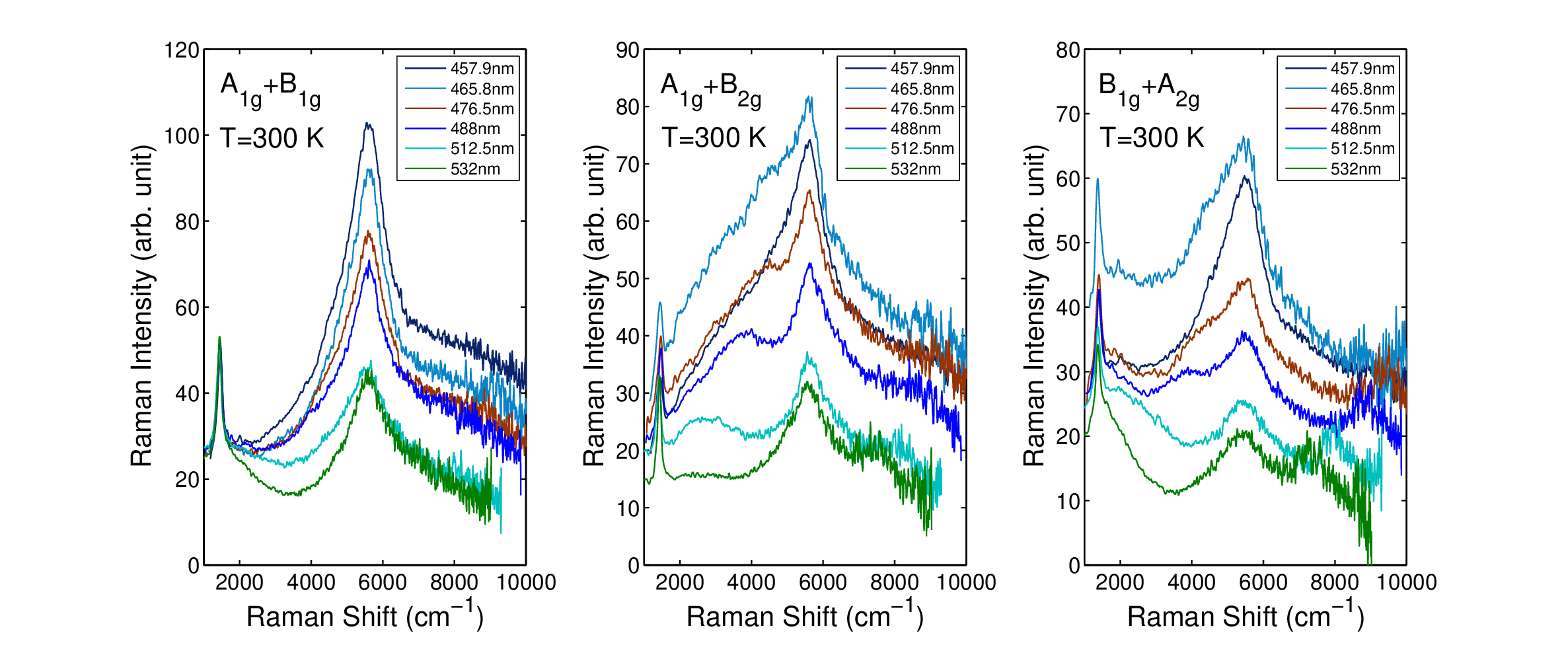} 


\caption{Raman spectra with different excitation energies in xx, x'x', and x'y' scattering configurations. The $A_{1g}$ and $B_{1g}$ peaks keep their positions at different laser energies, indicating that these peaks are real Raman signals. The spectra were normalized to the same power of incident lasers. \label{all_lines}}
\end{figure*}

The $A_{1g}$ and $B_{1g}$ features near 5600/5500 $cm^{-1}$ are enhanced with higher incident laser energy (shorter wavelength), while their positions remain the same, which is consistent with resonant Raman scattering (see Fig. \ref{all_lines}).


\begin{figure}[t!]
\includegraphics[width=0.5\textwidth]{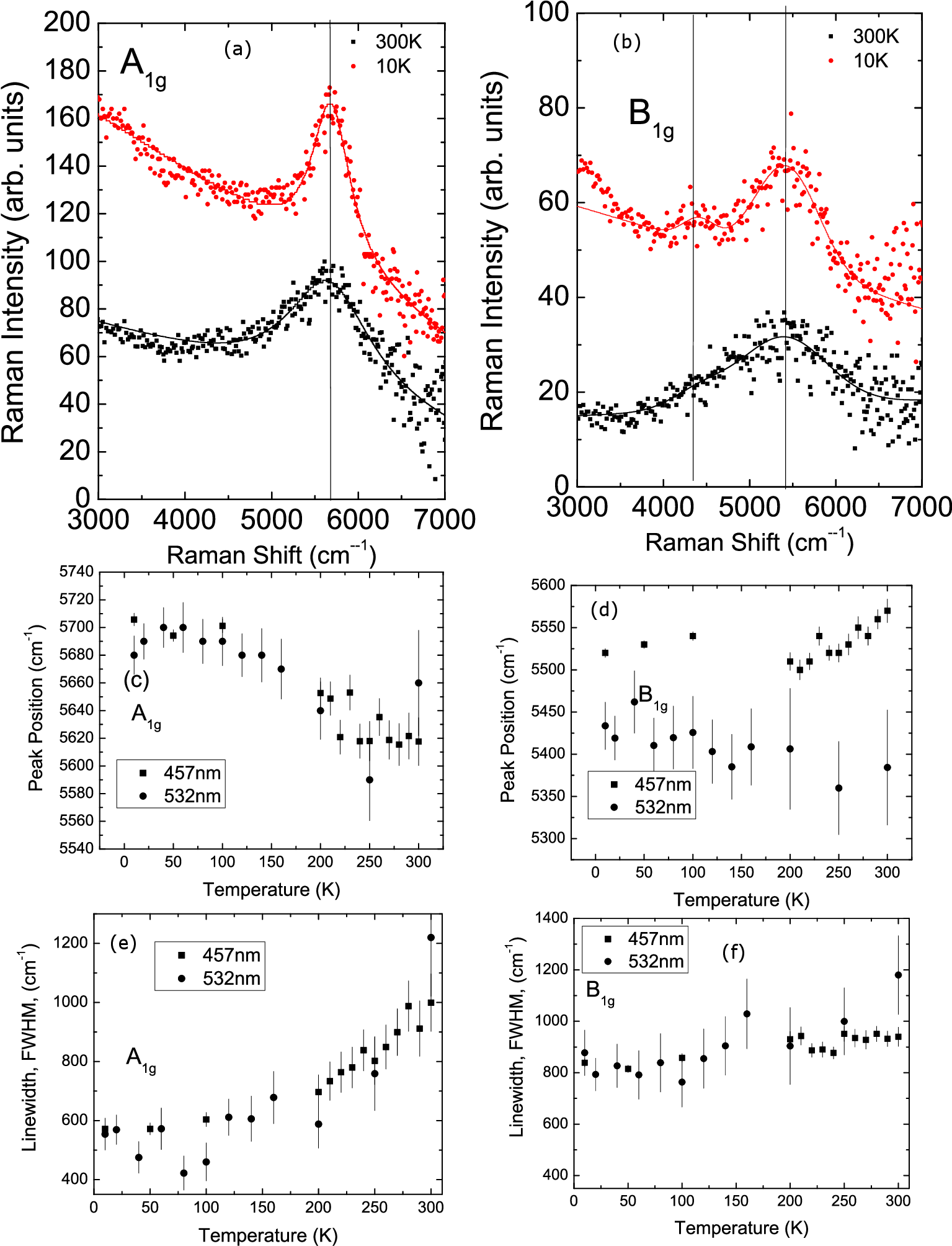} 
\caption{The comparison of Raman spectra at 10 K (red) and 300 K (black) in the (a) x'x' and (b) x'y' geometries with the laser at 532 nm. The solid lines indicate the peak positions. Temperature dependence of the peak positions and linewidth (FWHM) in the (c)(e) x'x' and (d)(f) x'y' geometries.
\label{temperature}}
\end{figure}

The $A_{1g}$ peak hardens on cooling from 5600 $cm^{-1}$ to 5700 $cm^{-1}$ saturating for $T<150$ K (Fig. \ref{temperature}). Its linewidth narrows on cooling from 1200 to 550 $cm^{-1}$ to below 200 K. The position of the $B_{1g}$ peak, is around 5450 $cm^{-1}$ at all temperatures, and its linewidth narrows from 1200 $cm^{-1}$ to 850 $cm^{-1}$ to below 200 K. At 300 K the $B_{1g}$ feature has a low-energy shoulder that appears as a distinct small peak at 4300  $cm^{-1}$ at low temperature. So the fits to the $B_{1g}$ spectrum include 2 components, a weak one at 4300 $cm^{-1}$ and a strong one at 5450 $cm^{-1}$.

\begin{figure}[t]
\includegraphics[width=0.5\textwidth]{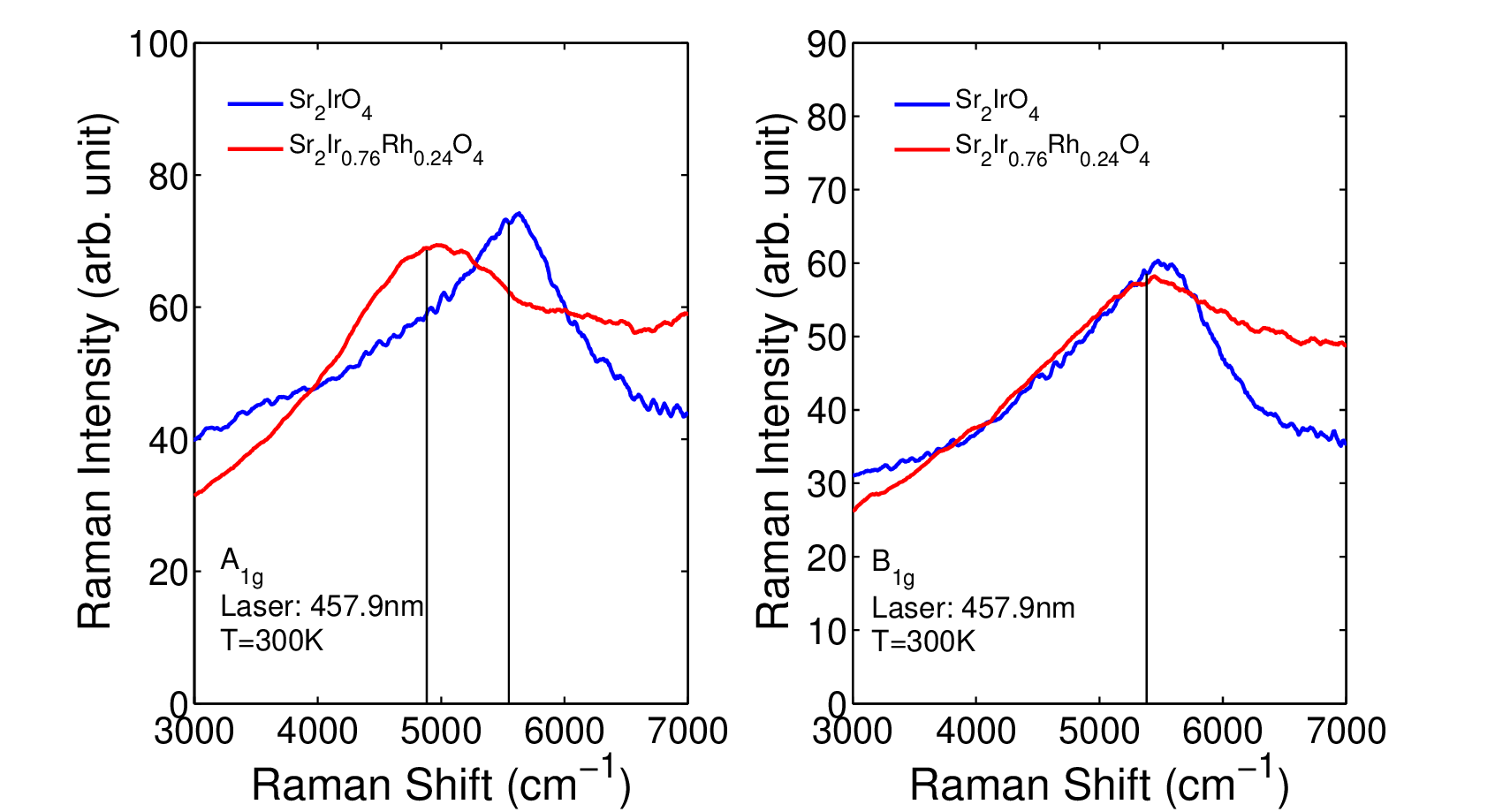} 
\caption{The comparison of Raman spectra in \ce{Sr2IrO4} (blue) and Sr$_2$Ir$_{0.76}$Rh$_{0.24}$O$_{4}$ (red) at room temperature with the laser at 457.9 nm.
\label{rh}}
\end{figure}

Substitution of 25$\%$ of Ir with Rh softens the $A_{1g}$ by 80 meV and the  $B_{1g}$ peak by 10 meV (Fig. \ref{rh}) and significantly broadens both peaks.

The peak energies are close to the insulating gap found in several experiments. Optical conductivity spectra exhibit two major peaks assigned to  transitions from the occupied $J_{\mathrm{eff}}$ = 1/2 (LHB) and 3/2 states to the unoccupied $J_{\mathrm{eff}}$ = 1/2 (UHB) state \cite{PhysRevB.80.195110,PhysRevLett.101.076402}. We do not expect an exact match to optical conductivity, because optically-active transitions are symmetry forbidden in Raman scattering and vice versa. Scanning tunneling microscopy (STM) studies show the onset of tunneling around 0.5 eV, with the separation between conductance peaks near 0.75 eV, which is near to what we measured \cite{PhysRevB.90.041102}. Our Raman features have a similar energy to the recent LDA calculations \cite{PhysRevB.87.245109,PhysRevLett.101.226402}. 

Previous work on the insulating compounds of the high $T_{c}$ cuprates revealed similar peaks in the $A_{2g}$ symmetry around 1.5-2 eV \cite{PhysRevB.51.6617,PhysRevB.53.886,PhysRevLett.71.3709}. Peaks at half the energy in the iridates are consistent with a much lower U, which in both cases is responsible for the insulating behavior. The d-d exciton that we observe is much sharper at low temperatures than the exciton peaks in most cuprates with the exception of the ones with the T' structure where several sharp peaks appear. Also, the peaks in cuprates are asymmetric and are located at the onset of a broad continuum of electronic scattering above the insulating gap. No such onset is seen in the iridates where the peaks are isolated and there is no broad band of scattering or higher energy peaks up to 1.2 eV.

\begin{figure}[t!]
\includegraphics[width=0.5\textwidth]{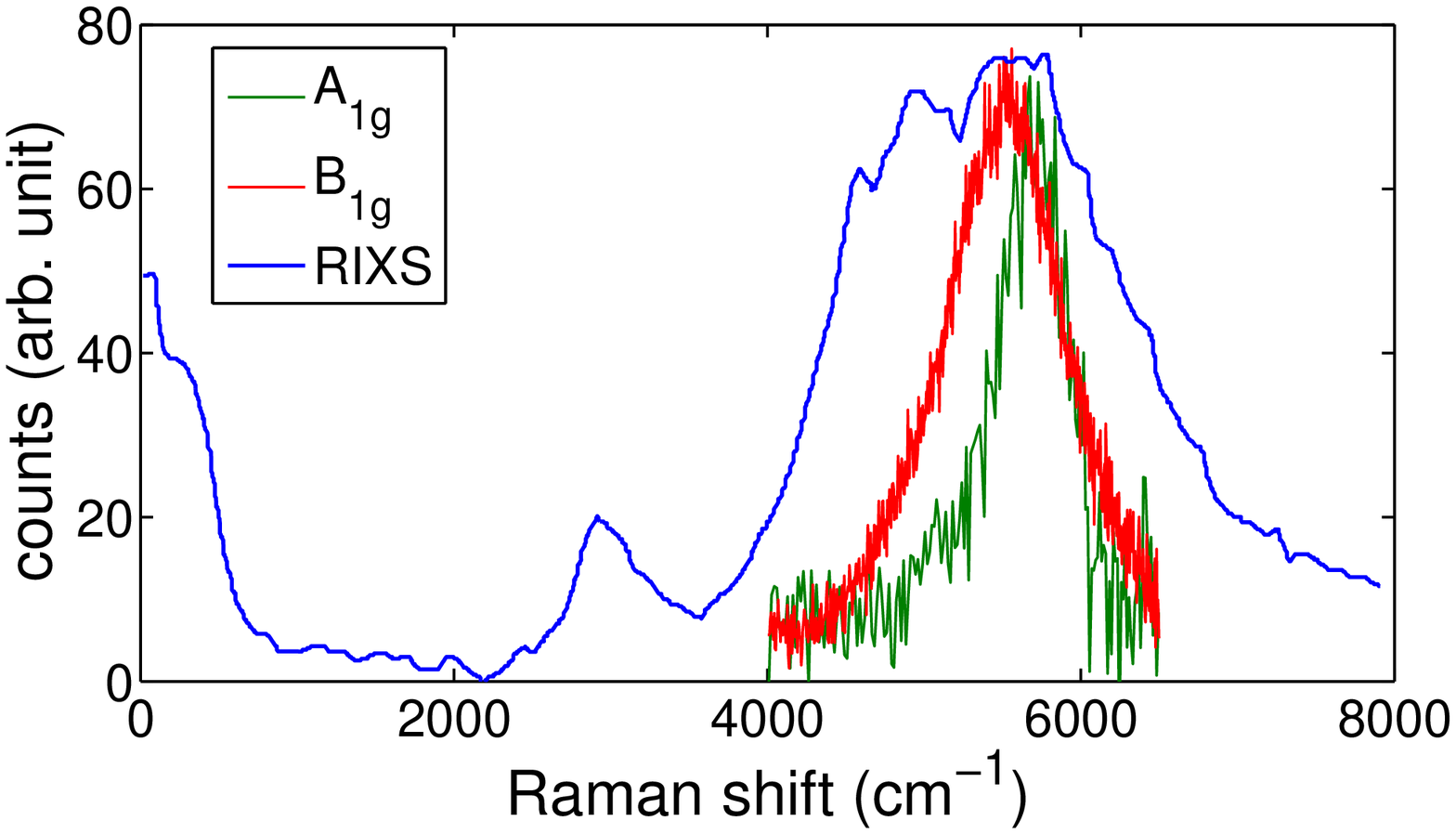} 
\caption{The comparison of Raman ($A_{1g}$ in green and $B_{1g}$ in red) and RIXS zone center data (blue) in Ref. \cite{nature_Kim}. A linear background is subtracted in the Raman spectra. The RIXS peaks around 4900 $cm^{-1}$ and 2900 $cm^{-1}$ are not observed in Raman scattering.
\label{RIXS_vs_Raman}}
\end{figure}

Our results are complementary to recently published RIXS data on the same compound, since RIXS observes the same excitations but with different selection rules and matrix elements. A direct comparison can be made between RIXS zone center data obtained with near normal incidence and our Raman data (Fig. \ref{RIXS_vs_Raman}). RIXS sees a band of scattering between 0.5 and 0.85 eV, which has a structure suggestive of several peaks. Our Raman results show that the structure seen in RIXS consists of the $A_{1g}$ and $B_{1g}$ Raman features and a third feature at lower energy not observed in Raman scattering. The latter feature is not seen by Raman either because it is symmetry forbidden, or because the Raman matrix element is negligibly small. We consider both possibilities in our analysis. 

\subsection{B. Selection rules for intra-site and inter-site transitions.}
First we compare these results to a simple calculation of what types of on-site and near neighbor hopping electronic transitions appear in what symmetry and then discuss how the interpretation of the RIXS data proposed in Ref. \cite{nature_Kim} needs to be reconsidered in light of the Raman results.

The Raman intensity is proportional to \cite{RevModPhys.79.175}:


\begin{equation}
I \propto \frac{1}{Z}  \sum_{I,F} \big| \langle F | R_{\mu \nu} | I \rangle \big|^2 e^{- E_I / k_B T} \delta(E_F - E_I - \hbar \omega) \text{,}
\end{equation}
where $I, F$ label energy eigenstates of the electronic system with energies
$E_I$, $E_F$, $\omega$ is the Raman shift, and $Z$ is the partition function.
$R_{\mu \nu}$ is the Raman tensor, with $\mu,\nu = x,y,z$ giving the direction of
linear polarization of scattered and incident light in our experiment,
respectively.  The electronic Raman cross section is typically dominated by the first two terms in perturbation theory [see Eq.~(13) of Ref.~\cite{RevModPhys.79.175}]. Fig. \ref{all_lines} shows that the intensity of the peaks of interest has a strong laser energy-dependence. Since the first order term does not depend on the laser energy , the second order term must dominate.  Thus we focus on this contribution, which is given by:
\begin{equation}
R_{\mu \nu} = p_{\mu} \frac{1}{E_I + \hbar \omega_I - H_{el} } p_{\nu} \text{,} \label{eqn:resonant}
\end{equation}
where $H_{el}$ is the electronic Hamiltonian, $p_{\mu}$ the electron momentum operator, and $\omega_I$ is the frequency of incident light.  [Note that we have dropped an additional non-resonant second-order term; see Eq.~(13) of Ref.~\cite{RevModPhys.79.175}.]



\begin{figure}[t]
	\begin{center}
		\includegraphics[width=\columnwidth]{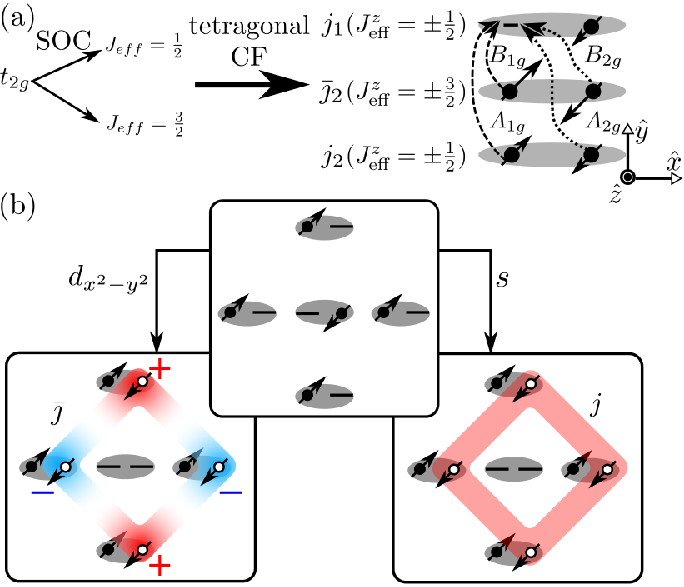}
	\end{center}
	\caption{Schematic of the on-site and inter-site transitions. (a) Local states and on-site transitions: The SOC and tetragonal CF split the $t_{2g}$ orbitals into three doublets labeled with $j_1$, $\barj_2$, and $j_2$. The non-pseudo-spin-flip electronic transition contributes to the $A_{1g}$ and $B_{1g}$ signal, and the pseudo-spin-flip process contributes the $A_{2g}$ and $B_{2g}$ signal.
    (b) Inter-site transitions: The ground state configuration is described in the top block. The electron can tunnel from the central site to the $s$-wave/$d_{x^2-y^2}$ cluster state which transforms identically to the on-site $j$/$\barj$ doublet respectively. The gray oval encloses the states on the same site.
    }
	\label{fig:cluster}
\end{figure}

In the presence of tetragonal crystal field (CF) and SOC, the $t_{2g}$ manifold splits into three Kramer doublets labeled with $j_1,j_2$ and $\barj_2$ (Fig. \ref{fig:cluster}(a)) \cite{PhysRevLett.109.157401,PhysRevB.84.020403,PhysRevLett.110.217212}. The $j$ ($J_{\mathrm{eff}}^{z}=\pm1/2$) and $\barj$ ($J_{\mathrm{eff}}^{z}=\pm3/2$) doublets transform differently under $D_{4h}$ symmetry and time reversal (See Appendix A). Note that $J_{\mathrm{eff}}^2$ is not a good quantum number under $D_{4h}$ CF.

We adopt a tight-binding description of the electronic states, with
$f^\dagger_{\br \alpha}$ creating an electron at the Ir lattice site $\br$ in
the local spin-orbital state $\alpha$. Here, the spin-orbital state
$\alpha\equiv(a,\sigma)$, where $a=j_1,j_2, \barj_2$ labels the local doublets, and $\sigma=1,2$ is the pseudo-spin.
We work in real-space. In our description of the Raman process, a photon is absorbed near a lattice site $\br$, with the resulting excited intermediate state propagating over some distance before emission of a photon near site $\br'$.  Far enough away from resonance, the intermediate state will propagate only over a short distance; this leads to the expansion
\begin{equation}
R_{\mu \nu} =  \sum_{\br} M^{\alpha \beta}_{0; \mu \nu}(\br) f^\dagger_{\br
\alpha} f^{\vphantom\dagger}_{\br \beta} 
+ M^{\alpha \beta}_{1; \mu \nu}(\br) c^\dagger_{\br
\alpha} f^{\vphantom\dagger}_{\br \beta} 
+ \cdots \text{.}
\label{RamanTensor}
\end{equation}
Here the first term represents on-site transitions, while the second term
describes inter-site processes, in which an electron moves from a site $\br$ to a cluster state created by $c^{\dagger}_{\br\alpha}$.  The cluster state is 
a linear superposition of spin-orbital states on the four Ir sites nearest to $\br$.
Sums over repeated indices
$\alpha, \beta$ are implied. Longer range terms have been dropped.


To interpret the experimental results, we assume significant local antiferromagnetic correlations to temperatures well above the Neel temperature ($T_N=240$ $\rm {K}$), focusing here on $T>T_N$, as the gross features of the Raman spectrum remain largely unchanged as $T$ is lowered through $T_N$. In the low-temperature antiferromagnetic state, the  moments lie in the $ab$-plane, with small canting out of the plane that we ignore \cite{Kim06032009,0953-8984-25-42-422202,PhysRevB.87.144405,PhysRevB.87.140406}.

\begin{table}
\begin{tabular}{c | c | c}
 & Non-pseudospin-flip & Pseudospin-flip  \\
 \hline
 $j_2 \to j_1$ & $x x$, $x' x'$ ($A_{1g}$) &  $x y$, $x' y'$ ($A_{2g}$) \\
 \hline
 $\barj_2 \to j_1$ & $x x$, $x' y'$ ($B_{1 g}$) & $x' x'$, $x y$ ($B_{2g}$) 
\end{tabular}
\caption{Polarization and symmetry of on-site Raman transitions with $D_{4h}$ site symmetry.  The row indicates the doublets between which the transition occurs, and the column indicates whether a pseudospin flip is involved.}\label{tab:pol}
\end{table}

We first consider an idealized situation, where the Ir-O-Ir bond angle is
180$^{\circ}$ and the site symmetry is $D_{4h}$.  We focus on on-site transitions within the $t_{2g}$ manifold, so  $j_2 \to j_1$ and $\barj_2 \to j_1$ are relevant  (the energy of the main peak is likely too large for purely magnetic $j_1 \to j_1$ transitions)(Fig. \ref{fig:cluster}(a)).  The pseudospin structure of each transition is described by the appropriate $2 \times 2$ block of the $6 \times 6$ matrix $M_{0 ; \mu\nu}(\br)$.  We always find this $2 \times 2$ matrix to be either proportional to the identity matrix (no pseudospin flip), or to the $\sigma^z$ Pauli matrix (pseudospin flip). 

For the on-site transitions, both non-pseudospin-flip and pseudospin-flip processes can occur (See Appendix A). The former should appear in $A_{1g}$ and $B_{1g}$ symmetry and the latter in $A_{2g}$ and $B_{2g}$ symmetry (Table ~\ref{tab:pol}). 


For the inter-site transitions, we consider hopping between $j_1$ doublets on neighboring sites.  An electron hops from site $\br$ into a parity-even cluster state, constructed by superposing $j_1$ doublets on the four neighboring sites (Fig. \ref{fig:cluster}(b)). 
There are two such Raman-active cluster states, one with $s$-wave symmetry and the other with $d_{x^2 - y^2}$ symmetry. In addition, there can be an infrared-active, Raman inactive transition to a p-wave cluster state. The $s$-wave/$d_{x^2 - y^2}$ cluster state transforms identically to the on-site $j$/$\barj$ doublet.  This means the $\br \to s$/$\br \to d_{x^2 - y^2}$ inter-site process has the same selection rules as the on-site $j_2 \to j_1$/$\barj_2 \to j_1$ transition respectively. Note that pseudospin-flip processes are forbidden by the combination of local antiferromagnetic correlations and the Pauli principle (see Fig.~\ref{fig:cluster}(b)).


So far we assumed $D_{4 h}$ site symmetry, but, in reality, the Ir-O-Ir bond angle is away from 180$^{\circ}$ by 22$^{\circ}$, lowering the site symmetry to $C_{4 h}$ (the point group remains $D_{4 h}$).  In this case, both $j_2 \to j_1$ and $\barj_2 \to j_1$ transitions (and corresponding inter-site transitions) may produce a non-pseudospin-flip contribution in $xy$ polarization.  That this is not seen suggests the site symmetry is effectively $D_{4h}$ to a good approximation.

\subsection{C. Discussion}
Raman scattering has different matrix elements from RIXS, and thus highlights different excitations. RIXS data are consistent with two dispersing modes whose spectral intensity can be controlled by the scattering angle. These have been assigned (in our notation) to excitons associated with  $j_2 \to j_1$ and $\barj_2 \to j_1$ intra-site transitions. However, these transitions should produce four distinct excitations, considering that the pseudospin flip and non-flip transitions should have different energies in the presence of magnetic-order. Here we propose that modes seen in RIXS as well as in our data can also originate from inter-site excitations whose energy would correspond to on-site repulsion. Thus more analysis and data are needed to understand the signal seen in RIXS.

Our Raman data narrows down the range of possible interpretations. It is clear that there are three sharply peaked distinct modes around 0.65 eV, not two (Two are Raman active but with different symmetries and one is either Raman inactive or its Raman matrix element is much smaller than for the other two)(Fig. \ref{RIXS_vs_Raman}). 

In the on-site transitions scenario this means that both pseudospin non-flip transitions are seen close to 0.7 eV with $j_2 \to j_1$/$\barj_2 \to j_1$ appearing in the $A_{1g}$/$B_{1g}$ symmetry respectively. This scenario necessitates that one or both corresponding pseudospin-flip transitions contribute to the RIXS peak at 0.6 eV and their Raman matrix element is so small that they are not seen in the XY-polarized Raman spectrum where they would appear. While this scenario cannot be ruled out without a better understanding of the Raman matrix elements, we would like to point out that the Raman data were taken over a wide range of laser energies, which always produced no signal in this scattering geometry. We will attempt to look for these excitations covering a wider range of both laser energies and Raman shifts in a future study. This scenario also implies that the splitting between the $j_2$ and $\barj_2$ levels is 30 meV, not 137 meV as proposed in Ref. \cite{nature_Kim}.

Another possibility is that the RIXS experiments reveal excitons associated with inter-site transitions. In this case three peaks come out naturally with transitions to the s-wave/d-wave cluster states making up the $A_{1g}$/$B_{1g}$ peaks and the transitions to the p-wave cluster making up the Raman-inactive peak at a lower energy. We note that pseudospin-flip transitions in this scenario are not allowed due to the Pauli exclusion principle.

The third possibility is that the three peaks come from some combination of inter-site and intra-site transitions. We think that this possibility is least likely, because it implies that the on-site repulsion energy should be very similar to the intra-site level splitting, which would be an unlikely coincidence.

We note that the simple picture above does not explain the differences in temperature and Rh-doping dependence between the $A_{1g}$ and $B_{1g}$  peaks,  although the broadening may be due to inhomogeneous doping in the sense that the electronic state is different near doped Rh atoms. More work is necessary to elucidate this issue. A more sophisticated calculation that would include band structure is necessary, because the two symmetries probe different k-space regions.

The narrow linewidth at 10 K of the two features is remarkable, especially the 50 meV linewidth of the $A_{1g}$ feature. In the band picture, zone center peaks originate from vertical inter band transitions, i.e their lineshape should reflect the distribution of the separations between valence and conduction bands throughout the Brillouin zone. In order to have very narrow peaks, the valence and conduction bands must be nearly parallel. DFT calculations do show nearly parallel bands \cite{PhysRevB.87.245109}, but not so as to produce such narrow peaks. Excitonic effects likely play a role in the final state as suggested in \cite{PhysRevLett.108.177003,nature_Kim}, which may be responsible for their small linewidth.

\section{IV. CONCLUSION}
To conclude, Raman results combined with recently published RIXS data reveal three sharp and closely spaced electronic excitations around 0.65 eV in \ce{Sr2IrO4}. Two of these appear in the Raman scattering spectra in the  $A_{1g}$ and $B_{1g}$ symmetries in \ce{Sr2IrO4}. These peaks originate from different electronic transitions as evidenced by their different temperature and Rh-doping dependence. In addition, the third peak at a somewhat lower energy has been reported in RIXS, but is not seen in the Raman spectra measured with visible light. We showed that several scenarios can describe these peaks on a purely qualitative level, but more work is necessary to provide a quantitative description.
\section{ACKNOWLEDGMENTS}
We thank A.V. Chubukov and T.P. Devereaux for discussions. Experimental work at the University of Colorado was supported by the National Science Foundation (NSF) under grant No. DMR-1410111. Theoretical work was supported by the U.S. Department of Energy (DOE), Office of Science, Basic Energy Sciences (BES) under Award No. DE-FG02-10ER46686 (Y.-P. H. and M.H.), and by Simons Foundation grant No. 305008 (M.H. sabbatical support). G.C. acknowledges NSF support via Grant No. DMR-1265162.

\appendix
\section{Appendix A: Supplementary information}

Here, we give details on how symmetry constrains the polarization dependence and pseudospin structure of the electronic Raman transitions considered in the main text.  We focus on on-site transitions; as stated in the main text, the inter-site transitions discussed are of the same symmetry as on-site transitions, and do not need to be considered separately here.  In addition, we show that the Raman tensor can indeed induce pseudospin-flip processes, even if $R_{\mu \nu}$ is assumed to act only on orbital (and not spin) degrees of freedom.

Beginning with Eq.~(3) of the main text, the objective is to use site symmetry and time reversal to constrain the matrix elements for on-site transitions, contained in the $6 \times 6$ matrix $M_{0; \mu \nu}$.  We focus on a single lattice site $\br$ and thus drop the site label from our analysis.  As discussed in the main text, we consider an idealized case of $D_{4h}$ site symmetry and only later consider breaking down to $C_{4h}$.  The analysis proceeds in the high-temperature phase, with no spontaneous symmetry breaking due to long-range magnetic order.

We focus on two $2 \times 2$ blocks of $M_{0; \mu \nu}$, one describing transitions from a $j$-doublet to another $j$-doublet $M^{j \to j}_{\mu \nu}$), and another describing transitions from a $j$-doublet to a $\barj$-doublet ($M^{j \to \barj}_{\mu \nu}$).  The on-site $j_2 \to j_1$ transition and the inter-site $\br \to s$ transition are both of $j \to j$ type, while the on-site $\barj_2 \to j_1$ and inter-site $\br \to d_{x^2 - y^2}$ transitions are of $j \to \barj$ type.  (The symmetry constraints on $j \to \barj$ and $\barj \to j$ transitions are the same.)

Ignoring inversion, which acts trivially on the electronic states of interest, $D_{4h}$ is generated by the operations $C_{4z}$ (four-fold rotation about the $z$-axis), $C_{2x}$ (two-fold rotation about the $x$-axis), and $C_{2xy}$ (two-fold rotation about the $(\hat{x}+\hat{y})$-axis).  It will also be useful to explicitly consider $C_{2z} = (C_{4z})^2$.  We consider the single-ion Hamiltonian obtained by projecting
spin-orbit coupling and $D_{4h}$ crystal field to the $t_{2g}$ manifold, which allows us to obtain wave functions for the electronic states of interest.  Using these wave functions, we find the following matrices representing the action of $D_{4h}$ symmetry on the $j$-doublets:
\begin{eqnarray}
C^j_{4z} &=& - \left( \begin{array}{cc} e^{- i \pi / 4} & 0 \\ 0 & e^{i \pi / 4} \end{array}\right) \\
C^j_{2z} &=& - i \sigma^z \\
C^j_{2x} &=& i \sigma^x \\
C^j_{2xy} &=&  \frac{-i}{\sqrt{2}} ( \sigma^x + \sigma^y ) \text{,}
\end{eqnarray}
where $\sigma^x, \sigma^y, \sigma^z$ are the usual $2 \times 2$ Pauli matrices.
For $\barj$-doublets we find:
\begin{eqnarray}
C^{\barj}_{4z} &=& \left( \begin{array}{cc} e^{- i \pi / 4} & 0 \\ 0 & e^{i \pi / 4} \end{array}\right) \\
C^{\barj}_{2z} &=& - i \sigma^z \\
C^{\barj}_{2x} &=& i \sigma^x \\
C^{\barj}_{2xy} &=&  \frac{i}{\sqrt{2}} ( \sigma^x + \sigma^y ) \text{.}
\end{eqnarray}
In both cases time reversal is given by
\begin{equation}
{\cal T} = i \sigma^y K \text{,}
\end{equation}
where $K$ is the complex conjugation operator.  We note that these forms only depend on the symmetry properties of the electronic states, which are expected to be captured accurately in our simple treatment.

Now we analyze the constraints on the matrix elements.  First, we consider the action of symmetry on $R_{\mu \nu}$, which has to agree with that on the corresponding matrix elements.  We only need to consider those operations that take a given component of $R_{\mu \nu}$ into itself (or minus itself):
\begin{eqnarray}
C_{2 z} : R_{x x} &\to& R_{x x} \label{eqn:site-symm-first} \\
C_{2 x} : R_{x x} &\to& R_{x x} \\
C_{2 z} : R_{x' x'} &\to& R_{x' x'} \\
C_{2 x y} : R_{x' x'} &\to& R_{x' x'} \\
C_{2 z} : R_{x y} &\to& R_{x y} \\
C_{2 x} : R_{x y} &\to& - R_{x y} \\
C_{2 z} : R_{x' y'} &\to& R_{x' y'} \\
C_{2 x y} : R_{x' y'} &\to& - R_{x' y'} \text{.} \label{eqn:site-symm-last}
\end{eqnarray}
In addition, $R_{\mu \nu}$ is invariant under time reversal.

Now we consider the matrix elements $M^{j \to j}_{\mu \nu}$ and $M^{j \to \barj}_{\mu \nu}$.
In each case, time reversal allows the matrices $1_{2 \times 2}$ and $i \sigma^\mu$  ($\mu = x,y,z$) to appear with arbitrary real coefficients, where $1_{2 \times 2}$ is the $2 \times 2$ identity matrix.  For example,
time reversal allows the form $M^{j \to j}_{x x} = a_0 \cdot 1_{2 \times 2} + a_{\mu} i \sigma^{\mu}$, without yet imposing any other symmetries.  Using all symmetries gives
\begin{eqnarray}
M^{j \to j}_{x x}, M^{j \to j}_{x' x'}, M^{j \to \barj}_{x x}, M^{j \to \barj}_{x' y'} &\propto& 1_{2 \times 2} \\
M^{j \to j}_{x y}, M^{j \to j}_{x' y'}, M^{j \to \barj}_{x' x'}, M^{j \to \barj}_{x y} &\propto& \sigma^z \text{.}
\end{eqnarray}
The information provided in Table~I of the main text follows from these results.  We note in particular that only pseudospin-flip transitions contribute in $xy$ polarization.


\emph{Effect of $C_{4h}$ site symmetry.}  The true Ir site symmetry is $C_{4h}$, which is generated by the operations $C_{4z}$ and inversion.  Focusing on effects of this lower symmetry in $xy$ polarization, we find that $M^{j \to j}_{xy}$ and $M^{j \to \barj}_{xy}$ are both allowed to have a non-pseudospin-flip contribution.  The fact that a peak is not seen in $xy$ polarization suggests that the breaking of $D_{4h} \to C_{4h}$ is a weak effect, at least for the electronic states probed by our Raman measurements.

\emph{Below Neel temperature.}  Below the Neel temperature, long-range magnetic order lowers the site symmetry.  Assuming the moments lie in the $xy$ plane and point along the $x'$ axis, the site symmetry is generated by the operations $C_{2 z} {\cal T}$ and $C_{2 x y}$ \cite{0953-8984-25-42-422202,PhysRevB.87.144405,PhysRevB.87.140406}. All components of $R_{\mu \nu}$ are left invariant by $C_{2 z} {\cal T}$.  This operation acts on both doublets as the matrix $C_{2 z} {\cal T} = - i \sigma^x K$, and this allows the matrices (with real coefficients) $1, \sigma^x, \sigma^y, i \sigma^z$ to appear.  Focusing again on $xy$ polarization, there are no further constraints on $M^{j \to j}_{xy}$ and $M^{j \to \barj}_{xy}$, so non-pseudospin-flip transitions are allowed to contribute in $xy$ polarization.  The absence of a peak indicates that Neel order is not strong enough compared to electronic energy scales to have a significant effect on the Raman transitions probed.

\emph{Raman scattering can flip the pseudospin.}  In the absence of spin-orbit coupling, the Raman tensor cannot induce spin-flip processes.  This follows from the presence of ${\rm SU}(2)$ spin rotation symmetry, and the fact that $R_{\mu \nu}$ commutes with ${\rm SU}(2)$ spin rotations.  In the present case, there is substantial spin-orbit coupling, and ${\rm SU}(2)$ spin symmetry is not present, opening the possibility of pseudospin flips in the Raman process. 

To assess whether $R_{\mu \nu}$ indeed contains pseudospin-flip transitions with significant amplitude, we make the conservative assumption that spin-orbit coupling only enters in the initial and final $t_{2g}$ states, ignoring spin-orbit coupling in the intermediate state.  As a result $R_{\mu \nu}$ commutes with ${\rm SU}(2)$ spin rotations, and spin-flip processes are forbidden.  However, we find pseudospin-flip processes are nonetheless Raman active.  To illustrate this point we focus on a single lattice site for simplicity and assume:
\begin{equation}
R_{\mu \nu} = f^\dagger_{A \sigma} {\cal R}^{\mu \nu}_{A B} f^{\phantom\dagger}_{B \sigma} \text{.}
\end{equation}
Here $A,B = yz, xz, xy$ labels the $t_{2g}$ orbital states,  $\sigma = \uparrow,\downarrow$ is electron spin, and the $3 \times 3$ matrix ${\cal R}^{\mu \nu}$ is constrained by site symmetry and time reversal. We are thus assuming that $R_{\mu \nu}$ acts only on the orbital degrees of freedom.  Spin-orbit coupling enters via the single-ion Hamiltonian, whose local energy eigenstates (the $j_1$, $j_2$ and $\barj_2$ doublets) have mixed spin and orbital character.  We determined the general symmetry-allowed form of ${\cal R}^{\mu \nu}$.  Then transforming the expression for $R_{\mu \nu}$ into the basis of spin-orbital energy eigenstates, we find that pseudospin-flip processes are fully allowed, even though  $R_{\mu \nu}$ contains no spin-flip terms.  This justifies our analysis of the Raman process including both transitions with and without pseudospin flips.

\bibliographystyle{apsrev4-1}
\bibliography{draft_reference}

\end{document}